\documentclass[reprint,amsmath,amssymb,aps,prl]{revtex4-1}
\usepackage[latin9]{inputenc}
\usepackage{verbatim}
\usepackage{amsmath}

\makeatletter
\usepackage{graphicx}
\usepackage{dcolumn}
\usepackage{bm}
\usepackage{tikz}
\usepackage{graphicx}
\usepackage{ragged2e} 
\usetikzlibrary{positioning}
\newcommand{\mi}{\mathrm{i}} 
\renewcommand{\vec}[1]{\mathbf{#1}}

\makeatother

\begin{document}
\title{Spontaneous time-reversal symmetry breaking without magnetism in a
$S=1$ chain}
\author{Shun-Chiao Chang}
\author{Pavan Hosur}
\affiliation{Texas Center for Superconductivity and Department of Physics, University
of Houston, Houston, Texas 77204, USA~\\
}
\begin{abstract}
States of matter that break time-reversal symmetry are invariably
associated with magnetism or circulating currents. Recently, one of
us proposed a phase, the directional scalar spin chiral order (DSSCO), as an exception:
it breaks time-reversal symmetry via chiral ordering of spins along a particular direction,
but is spin-rotation symmetric. In this work, we prove the existence
of this state via state-of-the-art density matrix renormalization group
(DMRG) analysis on a spin-1 chain with nearest-neighbor bilinear-biquadratic
interactions and additional third-neighbor ferromagnetic Heisenberg
exchange. Despite the large entanglement introduced by the third-neighbor coupling, we are able to access system sizes up to $L=918$ sites. We find first order phase transitions from the DSSCO
into the famous Haldane phase as well as a spin-quadrupolar phase where spin nematic
correlations dominate. In the Haldane phase, we propose and demonstrate
a method for detecting the topological edge states using DMRG that
could be useful for other topological phases too.

\end{abstract}
\pacs{Valid PACS appear here}

\maketitle

\section{\label{sec:introduction}Introduction}

Equilibrium states of matter that break time-reversal symmetry (TRS)
invariably contain a finite density of angular momentum, either spin
or orbital. Common examples such as magnets contain local
spin moments, while more complex ones include orbital moments, such
as loop current phases \cite{Varma1997,He2012}, anomalous Hall states
\cite{Haldane1988,Nagaosa2010}, and various chiral topological phases
\cite{Kalmeyer1987,Yao2007,Wen1989,Dhar2013,Zalatel2014,Kallin2009,Leggett1975,Osheroff1972,Osheroff1972a,Tsuruta2015}.
A property shared by these phases is that TRS is immediately restored when the moments melt. Thus, TRS-breaking is usually considered synonymous
with the formation of local moments, even though the latter also violate
spatial symmetries of the lattice.

 On the other hand, one of the authors
recently proposed an exception to this rule, namely, the directional
scalar spin chiral order (DSSCO) \cite{dssco}. In one-dimension,
the DSSCO can be thought as a state in which quantum fluctuations
have melted classical spin order in accordance with the Mermin-Wagner-Hohenberg-Coleman (MWHC)
theorem \cite{Mermin1966, Hohenberg1967, Coleman1973} and restored SU(2) spin-rotation symmetry
(SRS), but a vestigal scalar spin-chiral order captured by the
order parameter

\begin{equation}
\chi=\frac{1}{L}\sum_{i}^ {}\langle\bold S_{i}\cdot\bold S_{i+1}\times\bold S_{i+2}\rangle
\end{equation}
where $\bold S_{i}$ is the spin on the $i^{th}$ site, has survived.
Since $\bold S\to-\bold S$ under time-reversal, $\chi$ is an Ising
order parameter that breaks TRS, but preserves SRS. It is reminiscent
of some other phases that involve scalar spin chirality \cite{Grohol2005,Lee2013,CSL}.
The key difference is that the chirally correlated spins in all these examples lie
on the vertices of a triangle. Hence, they break enough spatial symmetries
to permit a moment perpendicular to its face, even if the on-site
moment vanishes. In contrast, the corresponding sites in the DSSCO are
collinear, so no such current is possible. Higher dimensional versions
of the DSSCO rely on thermal or disorder-driven fluctuations for the
restoration of SRS, with the latter proposed to
be pertinent to the long-standing problem of the pseudogap phase of
the cuprate superconductors, which show TRS-breaking in Kerr effect
measurements \cite{Xia2008,Spielman1992,Spielman1990,Karpetyan2014}
but no signs of magnetism in nuclear magnetic resonance \cite{Wu2015}.

Ref. \cite{dssco} presented the DSSCO as a phase that is allowed
by fundamental laws of quantum mechanics. However, it did not prove
its existence in a realistic model. Through large-scale  density matrix renormalization group (DMRG) analysis of a spin-1 chain, we fill this gap in knowledge by showing that the 
following spin-1 chain has the DSSCO as its ground state
in a wide regime of parameters:
\begin{equation}
H=\sum_{i}K\left[\cos\theta\left(\mathbf{S}_{i}\cdot\mathbf{S}_{i+1}\right)^{2}+\sin\theta\mathbf{S}_{i}\cdot\mathbf{S}_{i+1}\right]-J\mathbf{S}_{i}\cdot\mathbf{S}_{i+3}\label{eq:Hamiltonian}
\end{equation}
Here, $J,K>0$ and $\theta\in[0,\pi/2]$ parametrizes the relative
strengths of the bilinear and the biquadratic nearest-neighbor couplings.
When $J=0$, $H$ reduces to the bilinear-biquadratic model
that was studied in Ref. \cite{bibqmodel} and shown to realize a
quasi-long-range ordered spin-quadrupolar (SQ) phase for $0<\theta<\pi/4$
and the Haldane phase $\theta>\pi/4$, separated by a Berezinskii-Kosterlitz-Thouless
type phase transition at $\theta=\pi/4$ \cite{BKT}. We explore the effects of non-zero $J$ on this model and find that the SQ is driven into the DSSCO, either directly or via intermediate Haldane and disordered phases, while the Haldane phase simply disorders at finite $J$ for most values of $\theta$. In the $J\to\infty$ limit, $H$ reduces to three copies of a Heisenberg ferromagnet, while finite $J$ introduces quantum fluctuations that melt the ferromagnet in accordance with the MWHC theorem \cite{Mermin1966, Hohenberg1967, Coleman1973}. The full phase diagram is shown in Fig. \ref{fig:phasedia}.

The emergence of the DSSCO as a ground state of $H$ can be anticipated heuristically as follows. In the classical limit, $S\to\infty$, the biquadratic term $K\cos\theta$ dominates and forces adjacent spins to be mutually orthogonal. At $\theta=0$, the remaining ferromagnetic coupling $J$ favors parallel third neighbors, resulting in two degenerate ground state manifolds $R|x,y,z,x,y,z,\dots\rangle$ and $R|-x,-y,-z,-x,-y,-z,\dots\rangle$ that are related by time-reversal and have opposite expectation values of $\chi$. Above, $\pm x$ at the $i^{th}$ position in the ket denotes a state with spin at the $i^{th}$ site maximally polarized along $\pm x$ and $R\in SU(2)$ represents an arbitrary global spin rotation. The classical ground state is then randomly chosen from these manifolds, thus breaking TRS and SRS spontaneously. For finite $S$, quantum fluctuations produce smooth deformations in the magnetization texture or gapless spin waves. In one dimension, these fluctuations are strong enough to melt the underlying spin order and restore SRS \cite{Mermin1966, Hohenberg1967, Coleman1973}. However, smooth deformations cannot change the chirality of the ground state, thus yielding the DSSCO. In this work, we find that the ground state at $\theta=0$ is the boundary between ferromagnetic phase and DSSCO. A non-zero $\theta$ is needed to stabilize the DSSCO. However, a large $\theta$ again destabilizes it in favor of the Haldane or the disordered phase.

\section{\label{sec:numerics}Numerical Procedure}

We carry out state-of-the-art DMRG calculations using the ITensor library developed by Stoudenmire and White \cite{itensor}. We perform up to 215 sweeps with a final
maximum bond dimension of $m=800$, which restricts the truncation
error to below $10^{-6}$. We are able to access system sizes up to $L=918$ despite our model containing a third-nearest-neighbor interaction. In comparison, DMRG calculations on the simpler bilinear-biquadratic spin-1 chain which has only nearest-neighbor and next-nearest-neighbor terms can  only reach $L=300$ sites \cite{trimerization}. Such a dramatic improvement in the performance results from using a pinning field on open chains to diagnose the phases of interest. We elaborate on this technique below.

Naively, ordering is captured by the unbiased correlation function:
\begin{align}
m=\lim\limits _{L\to\infty}\sqrt{\frac{1}{L}\sum_{i}e^{\mi qi}\left\langle A_{1}A_{i}\right\rangle _{H}}\label{eq:1stmeth}
\end{align}
where $A_{i}$ is an operator that corresponds to the order parameter on the $i^{th}$ site and $H$ is the Hamiltonian whose ground state
the expectation value is computed in. In the current problem, we consider
three choices of $A_i$: spin $S_{i}^{z}$, quadrupole $Q_{i}^{zz}$
and $\chi_{i}=\vec{S}_{i}\cdot\vec{S}_{i+1}\times\vec{S}_{i+2}$.
A finite value of $m$ signals long-ranged order and spontaneous
breaking of symmetry. 
However, this approach requires very large system sizes and high precision
to obtain reliable results, since it computes the square of the local
order parameter, which can be a very small, especially close to a
phase boundary. This issue can be circumvented by adding a training
field with an appropriate Fourier component, $H'=h\sum_{i}e^{\mi qi}A_{i}$,
to $H$ and computing
\begin{equation}
m=\lim\limits _{h\to0}\lim\limits _{L\to\infty}\frac{1}{L}\sum_{i}e^{\mi qi}\left\langle A_{i}\right\rangle _{H+h\sum_{i}e^{\mi qi}A_{i}^{z}}\label{eq:2ndmeth}
\end{equation}
The ordering of limits is crucial: one first has to take the thermodynamic
limit and then the limit of vanishing training field $h\rightarrow0$.
Such an approach was used, for instance, in Ref. \cite{pinning15}.

We go a step further and consider a \emph{local} field $H''=h_{1}A_{1}$
localized on the first site (or first three sites when $A_{i}=\chi_{i}$).
This trick lifts the burden of taking $h\rightarrow0$ numerically.
In fact, we can make $h_{1}$ strong enough to saturate the order
at the first site \cite{pinning10}. Then, long-range order is captured
by
\begin{equation}
m=\lim\limits _{i\to\infty}\lim\limits _{L\to\infty}e^{\mi qi}\left\langle A_{i}^{z}\right\rangle _{H+h_{1}A_{1}}\label{eq:pinfield}
\end{equation}
That is, one first has to take the thermodynamic limit and then take
the distance from the pinning center to infinity. This approach
has been shown to be less sensitive to finite-size effects of the
order parameter than the other two methods \cite{pinning}.
In following sections, we will use this method to diagnose the spin
and SQ orders. Applying it to the DSSCO, however, causes
the code to get stuck in metastable states with fractionalized Ising
domain wall excitations \cite{isingdomainwall}.
Therefore, we use Eq. (\ref{eq:2ndmeth}) for DSSCO with $q=0$ since
a uniform field destabilizes the domain walls and helps find the true
ground state.

Although the Hamiltonian has SRS and thus commutes with $S_{total}^z = \sum_i S^z_i$, we found that implementing DMRG separately within each $S_{total}^z$ subspace resulted in significantly slower or sometimes, no convergence. We speculate that this may be because fixing $S_{total}^z$ to a non-zero value $N$ amounts to an interaction $H_U = U(S_{total}^z-N)^2 = U\sum_{i,j}S^z_iS^z_j + 2NU\sum_iS_i^z + \textrm{const.}, U\to\infty$. This contains coupling between spins that are far apart, which would tend to slow down the DMRG calculation. Luckily, ordinary gapped phases have short-ranged spin correlations, so the slowdown and the net effect of this fixing $S^z_{total}$ is to speed up the procedure by reducing the size of the Hilbert space. In the current problem, however, the range of spin-spin correlations is only limited by the MWHC theorem and hence is extremely large (as we also show below). Consequently, the speed-up because of a smaller Hilbert space cannot offset the slowdown due to the long-range coupling in $H_U$. Thus, in our implementation, we allow the program to explore different values of $S_{total}^z$ while searching for the ground state.

\section{\label{sec:phase-diagram}Results: PHASE DIAGRAM}

\begin{figure}
\begin{tikzpicture}[black]
	
	\node (img1)  {\includegraphics[height=6cm, width=8cm]{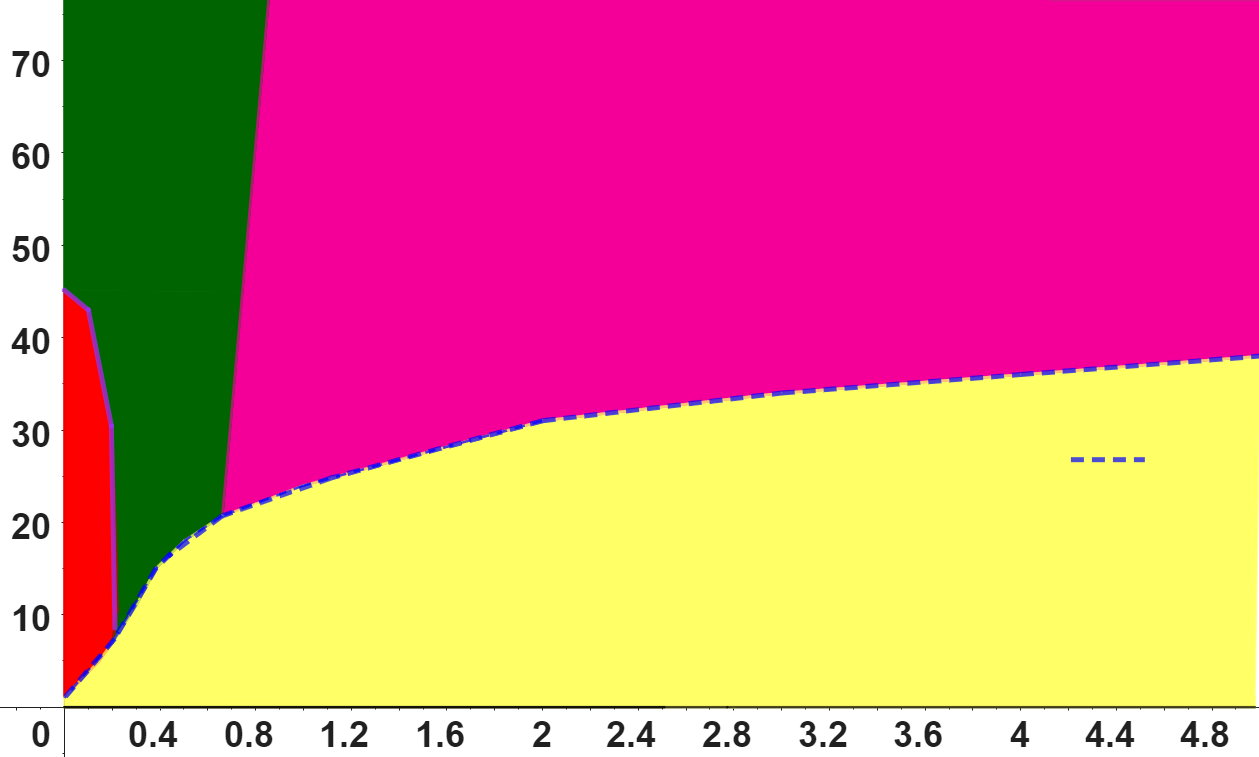}};
	\node[below=of img1, node distance=0cm, yshift=1cm,font=\color{black}] {$J/K$};
	\node[left=of img1, node distance=0cm, rotate=90, anchor=center,yshift=-0.9cm,font=\color{black}] {\large $\theta$(degree)};
     \node at (0.0,-1.0) {\large \color{purple} DSSCO};
     \node at (0.0,1.0) {\color{black}Disorder};
     \node at (-3.0,1.5) {\color{white}Haldane};
     \node at (-3.45,-1.0) {\tiny SQ};
     \node at (1.8,-0.6) {1st order line};
     \node at (2.0,-1.0) {};		
	
	\end{tikzpicture} \caption{The ground state phase diagram of the spin-1 Hamiltonian in (\ref{eq:Hamiltonian}). The DSSCO, Haldane, SQ and disordered phases are identified. Dotted lines denote first-order phase transitions between the DSSCO and the other three phases.}
\label{fig:phasedia} 
\end{figure}
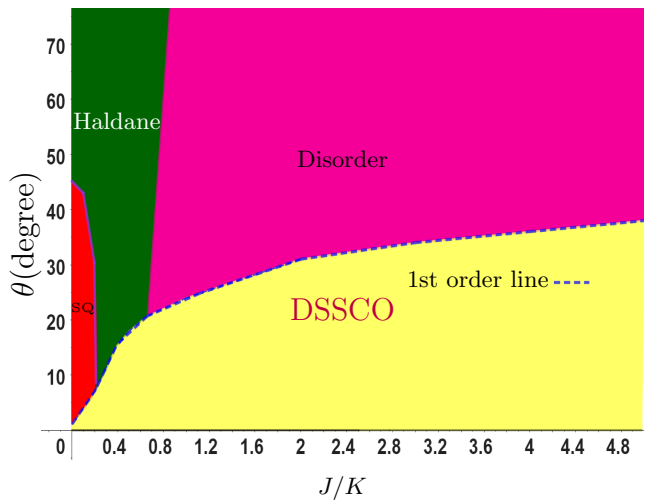

Fig. \ref{fig:phasedia} summarizes the phase diagram of the Hamiltonian
(\ref{eq:Hamiltonian}) obtained by DMRG. We reproduced known results
on the bilinear-biquadratic model \cite{bibqmodel} for the SQ and the Haldane phases on the $J=0$ axis. For small negative $\theta$ and $J=0$ the ground state is known to be ferromagnetic \cite{bibqmodel}. Unsurprisingly, we found (but do not show in Fig. \ref{fig:phasedia}) that the ferromagnet survives non-zero $J$. Interestingly, all the phases share boundaries with the DSSCO, which is the primary focus of this work.

\subsection{\label{sec:DSSCO}DSSCO Phase}

\begin{figure}
\begin{tikzpicture}[black]
	
	\node (img1)  {\includegraphics[height=5cm, width=8cm]{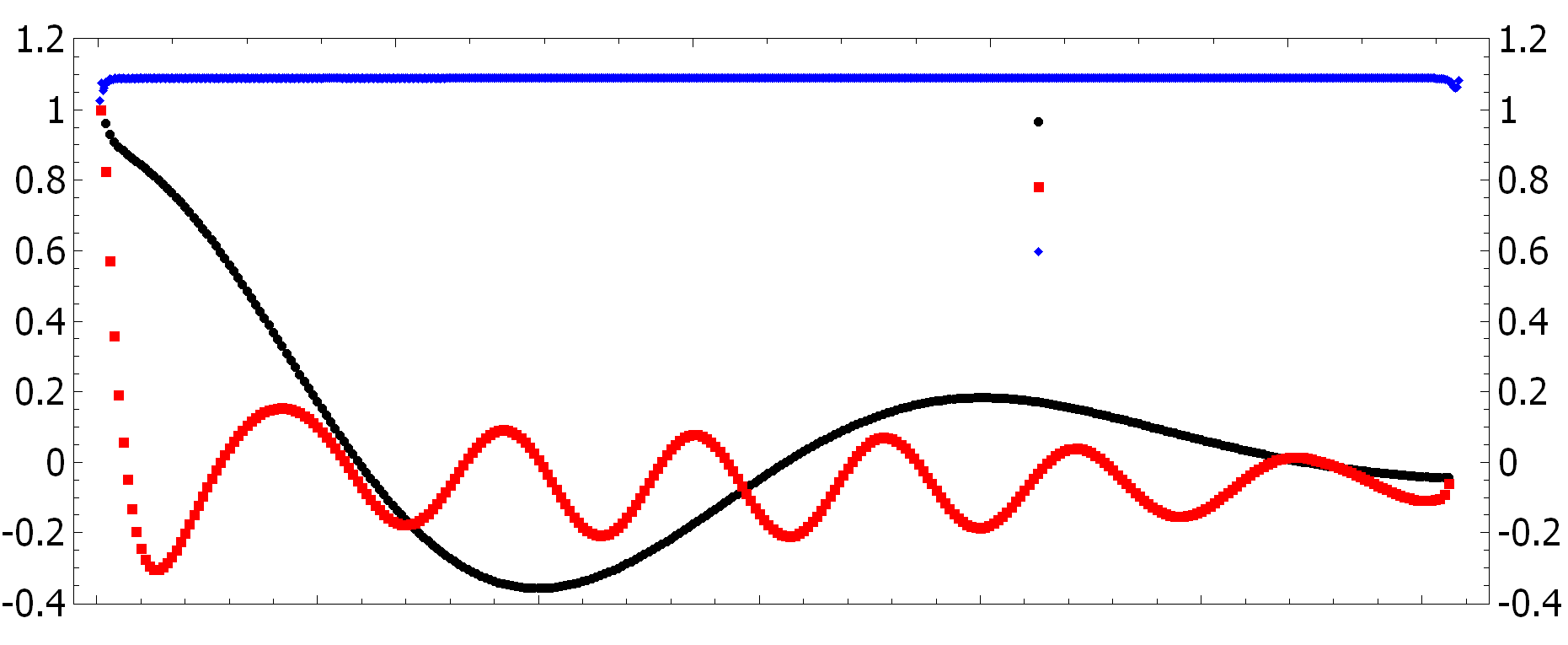}};
	\node at (2.5,1.5) {$\left\langle\ S^z_{3i-2}\right\rangle_{H+h_1 S^z_1}$};
	\node at (2.5,1.0) {$\tiny \langle\ Q^{zz}_{3i-2}\rangle_{H+q_1 Q^{zz}_1}$};
	\node at (2.5,0.5) {$\left\langle\ \chi_i\right\rangle_{H+\lambda\sum_{i}^{}\chi_i}$};
	\node at (-3.5,-2.4) {$0$};
	\node at (-2.4,-2.4) {$150$};
	\node at (-1.2,-2.4) {$300$};
	\node at (-0.1,-2.4) {$450$};
	\node at (1.0,-2.4) {$600$};
	\node at (2.15,-2.4) {$750$};
	\node at (3.25,-2.4) {$900$};
	\node[below=of img1, node distance=0cm, yshift=1cm,font=\color{black}] {\large $i$};
	\node[left=of img1, node distance=0cm, rotate=90, anchor=center,yshift=-1.0cm,font=\color{black}] {};
		
	\end{tikzpicture} \caption{Order parameters in the DSSCO phase measured by applying appropriate training fields. A weak uniform training chiral field $\lambda=0.01$ is used to probe the chiral order, whereas spin- and SQ correlations are probed by applying strong training fields $h_{1}=q_{1}=50$ at the first site. The spin and SQ orders decay to zero over long distances, which suggests the absence of long-range order in these variables. Data shown is for $L=918$ at $J=1$ and $\theta=20^{\circ}$.\label{fig:ordersite}}
\begin{tikzpicture}[black]
	
	\node (img1)  {\includegraphics[height=5cm, width=8cm]{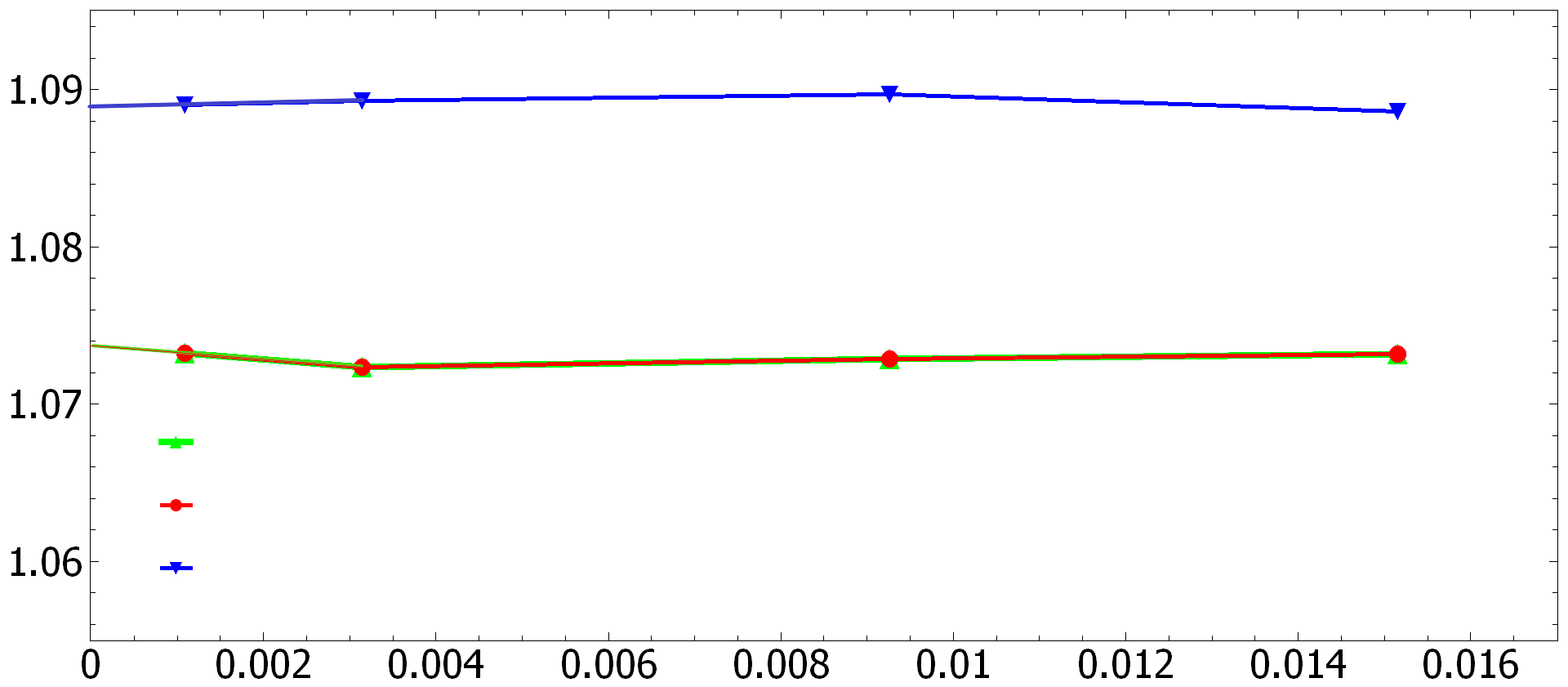}};

	\node[below=of img1, node distance=0cm, yshift=1cm,font=\color{black}] {$1/L$};
	\node[left=of img1, node distance=0cm, rotate=90, anchor=center,yshift=-1cm,font=\color{black}] {$\chi$};
	\node at (-2.2,-0.7) {$\lambda=10^{-4}$};
	\node at (-2.20,-1.2) {$\lambda=10^{-6}$};
	\node at (-2.20,-1.7) {$\lambda=10^{-2}$};
		
	\end{tikzpicture} \caption{Finite-size scaling of $\chi$ for multiple weak
uniform training chiral fields at $J=1$ and $\theta=20^{\circ}$. Clearly, $\chi$ survives as the training field is switched off, which indicates the formation of an ordered phase, namely, the DSSCO, via spontaneous symmetry breaking. \label{fig:fss}}
\end{figure}

The most exciting feature of the phase diagram is the DSSCO, which
breaks TRS but preserves SRS. We show numerical evidence for this phase in Figs. \ref{fig:ordersite} and \ref{fig:fss}. 

Fig. \ref{fig:ordersite} shows that
pinning the chirality of the first three sites induces chiral ordering
of $O(1)$ magnitude throughout the chain. Moreover, as shown in Fig.
\ref{fig:fss}, it robustly survives finite size scaling to the thermodynamic
limit, $L\to\infty$, even as the pinning field $\lambda$ is tuned
down. In contrast, Fig. \ref{fig:ordersite} shows that spin and SQ
orders decay to zero despite pinning their values on the first site.
Note, $\left\langle \mathbf{S}_{i}\right\rangle $ and $\left\langle \hat{\mathbf{Q}}_{i}\right\rangle $
have been shown on every third site. This is because the classical magnetic order that the DSSCO emerges from induces $q=2\pi/3$
oscillations in them that are not the subject of our interest; we
are interested in the amplitude of these oscillations only. %
{} Thus, in accordance with the MWHC theorem \cite{Mermin1966, Hohenberg1967, Coleman1973} which allows (forbids)
discrete (continuous) symmetry breaking in one-dimension, the DSSCO
breaks TRS and shows long-range order while spin and spin-quadrupoles
only show short-range order, since their order parameters break continuous SRS. Outside the
DSSCO, $\chi$ vanishes in the Haldane, SQ and disordered phases.

In the following, we will discuss spin order in the Haldane phase, where the bulk is naively disordered but a hidden order exists between the edges, as well as the SQ phase where either quasi-long range order or disorder exists, but SQ correlations dominate and have a large correlation length.

\subsection{\label{sec:Haldane}Pinning the Haldane Phase}

\begin{figure*}
\begin{tikzpicture}[black]
	\node (img1)  {\includegraphics[height=7cm, width=17cm]{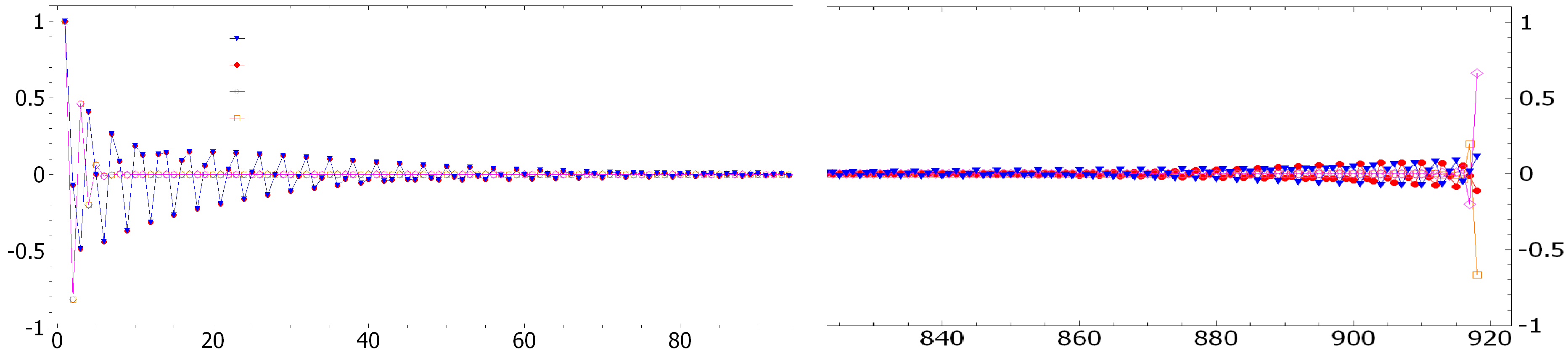}};
	\node[below=of img1, node distance=0cm,yshift=1cm,font=\color{black}] {\huge $i$};
	\node[below=of img1, node distance=0cm,xshift=0.1cm,yshift=2.0cm,font=\color{black}] {\huge $\wr$};
	\node[below=of img1, node distance=0cm,xshift=0.5cm,yshift=2.0cm,font=\color{black}] {\huge $\wr$};
	\node[below=of img1, node distance=0cm,xshift=0.1cm,yshift=5.1cm,font=\color{black}] {\huge $\wr$};
	\node[below=of img1, node distance=0cm,xshift=0.5cm,yshift=5.1cm,font=\color{black}] {\huge $\wr$};
	\node[below=of img1, node distance=0cm,xshift=0.1cm,yshift=8.4cm,font=\color{black}] {\huge $\wr$};
	\node[below=of img1, node distance=0cm,xshift=0.5cm,yshift=8.4cm,font=\color{black}] {\huge $\wr$};
	\node[left=of img1, node distance=0cm, rotate=90, anchor=center,yshift=-0.7cm,font=\color{black}] {\large $\left\langle\ S^z_i\right\rangle$};
	\node [label={[xshift=0.0cm, yshift=3.0cm]}] {};	
	\node at (-4,2.7) { Haldane $\lvert \uparrow_1 \uparrow_L \rangle$};
	\node at (-4,2.1) { Haldane $\lvert \uparrow_1 \downarrow_L \rangle$};
	\node at (-4,1.6) { AKLT $\lvert \uparrow_1 \uparrow_L \rangle$};
	\node at (-4,1.1) { AKLT $\lvert \uparrow_1 \downarrow_L \rangle$};
	
	\node (hlu) at (5.5,1.0) { Haldane $\lvert \uparrow_1 \uparrow_L \rangle$};
	\node (hld) at (5.5,-1.0) { Haldane $\lvert \uparrow_1 \downarrow_L \rangle$};
	\node (alu) at (5.5,2.0) { AKLT $\lvert \uparrow_1 \uparrow_L \rangle$};
	\node (ald) at (5.5,-2.0) { AKLT $\lvert \uparrow_1 \downarrow_L \rangle$};
	
	\node (Hu) at (-5.2,2.7) { };
	\node (Hd) at (-5.2,2.1) { };
	\node (Au) at (-5.2,1.6) { };
	\node (Ad) at (-5.2,1.1) { };
	\node (1st) at (-7.6,3.1) {};
	
	\node (hlud) at (7.5,0.4) {};
	\node (hldd) at (7.5,-0.3) {};
	\node (alud) at (7.5,2.0) {};
	\node (aldd) at (7.5,-2.0) {};
	\draw[->,dashed]
	(Hu) edge (1st) (Hd) edge (1st) (Au) edge (1st) (Ad) edge (1st);
	\draw[->,dashed]
	(hlu) edge (hlud) (hld) edge (hldd) (alu) edge (alud) (ald) edge (aldd);
	\end{tikzpicture} \caption{Edge states in the Haldane phase. Two degenerate ground states for Haldane phases at $\theta=15^{\circ}$,
$J/K=0.3$ and for AKLT point $\theta = 71.5^\circ, J=0$ are shown and the spins on the edges are marked by dashed arrows. The moment at the first site is saturated
by a local training field, and $\langle S_i^z\rangle$ is measured at all the sites. For each set of parameters $(\theta,J/K)$, $\langle S_i^z\rangle$ on the right edge is large while $\langle S_i^z\rangle$ in the bulk is extremely small. Note, bulk sites between $i\approx 100$ and $i\approx 820$ are not shown to highlight the edge states.}
\label{hald} 
\end{figure*}
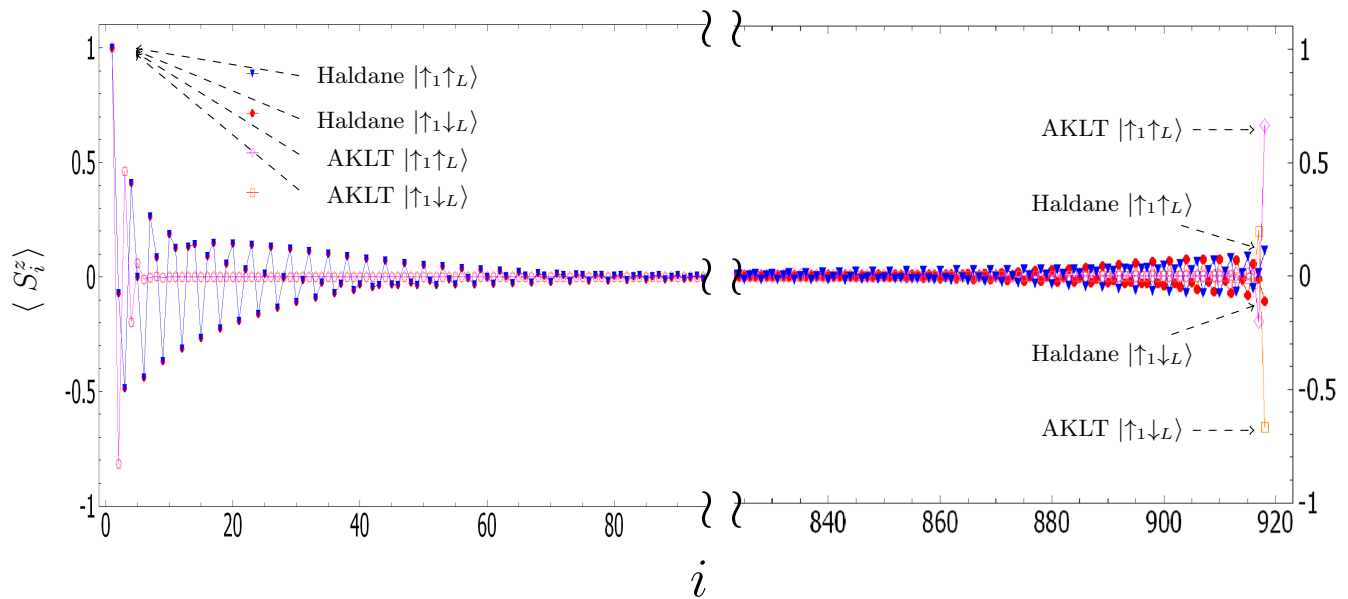
The Haldane phase is one of the simplest examples of a symmetry-protected topological phase \cite{1Hald1983,2Hald1983,Pollmann2012}. Its simplest realization is in the antiferromagnetic Heisenberg model, which is the $\theta=\pi/2, J=0$ limit of $H$, while the point $\theta=\tan^{-1}3\approx71.5^\circ, J=0$ is in the same phase and corresponds to the exactly soluble Affleck-Kennedy-Lieb-Tasaki (AKLT) point \cite{AKLT1987}. The Haldane phase has no local order parameter; instead, it can be characterized by a non-local string order parameter \cite{MPSstring2008, haldwhite} that captures entanglement between states on opposite ends of the chain. In particular,
the ground state in the Haldane phase is fourfold degenerate on an infinite open chain.
The degeneracy stems from the two effective spin-1/2s, one exponentially
localized at each end \cite{haldwhite}. For a finite chain, the states
at opposite ends hybridize, resulting in a unique singlet ground state: $\frac{1}{\sqrt{2}}(\lvert\uparrow_{1}\downarrow_{L}\rangle-\lvert\downarrow_{1}\uparrow_{L}\rangle)$,
and a threefold-degenerate triplet of excited states: $\lvert\uparrow_{1}\uparrow_{L}\rangle$,
$\frac{1}{\sqrt{2}}(\lvert\uparrow_{1}\downarrow_{L}\rangle+\lvert\downarrow_{1}\uparrow_{L}\rangle)$,
and $\lvert\downarrow_{1}\downarrow_{L}\rangle$. 
In the thermodynamic limit, the singlet and triplet sectors become exactly degenerate.
Therefore, the edge states can be detected by directly computing spin-spin correlation function in the ground state using (\ref{eq:1stmeth}), which is equivalent to calculating the string order. The correlation is non-trivial between the opposite ends of the chain, but vanishes between an edge site and a bulk site.

We expect the Haldane phase to occur in our model as well in a region of phase space around the Heisenberg and AKLT points. However, the third-neighbor interactions increase the ground state entanglement, which drastically increases the cost of computing the nonlocal order. We therefore adopt  an alternate strategy to detect the edge states that not only avoids measuring
the non-local order but also reduces the entanglement of our ground state. We apply a spin-pinning field on the first site,
which reduces the fourfold degenerate space to two doubly degenerate
subspaces, $\left(|\uparrow_{1}\uparrow_{L}\rangle,|\uparrow_{1}\downarrow_{L}\rangle\right)$
and $\left(|\downarrow_{1}\uparrow_{L}\rangle,|\downarrow_{1}\downarrow_{L}\rangle\right)$,
since the pinning field favors (disfavors) states with spin at the
first site parallel (anti-parallel) to the field. Fig. \ref{hald}
shows signatures of the edge states in the Haldane phase at $\theta=15^{\circ}$,
$J/K=0.3$, and at the AKLT point \cite{AKLT1987}. The exactly
soluble AKLT point shows sharp spin moments at the edges, whereas the
moments elsewhere in the Haldane phase decay exponentially into the
bulk. 

\subsection{\label{sec:spin-quadrupole}Spin-Quadrupole phase}

\begin{figure}
\begin{tikzpicture}[black]
\node (img1)  {\includegraphics[height=5cm, width=8.5cm]{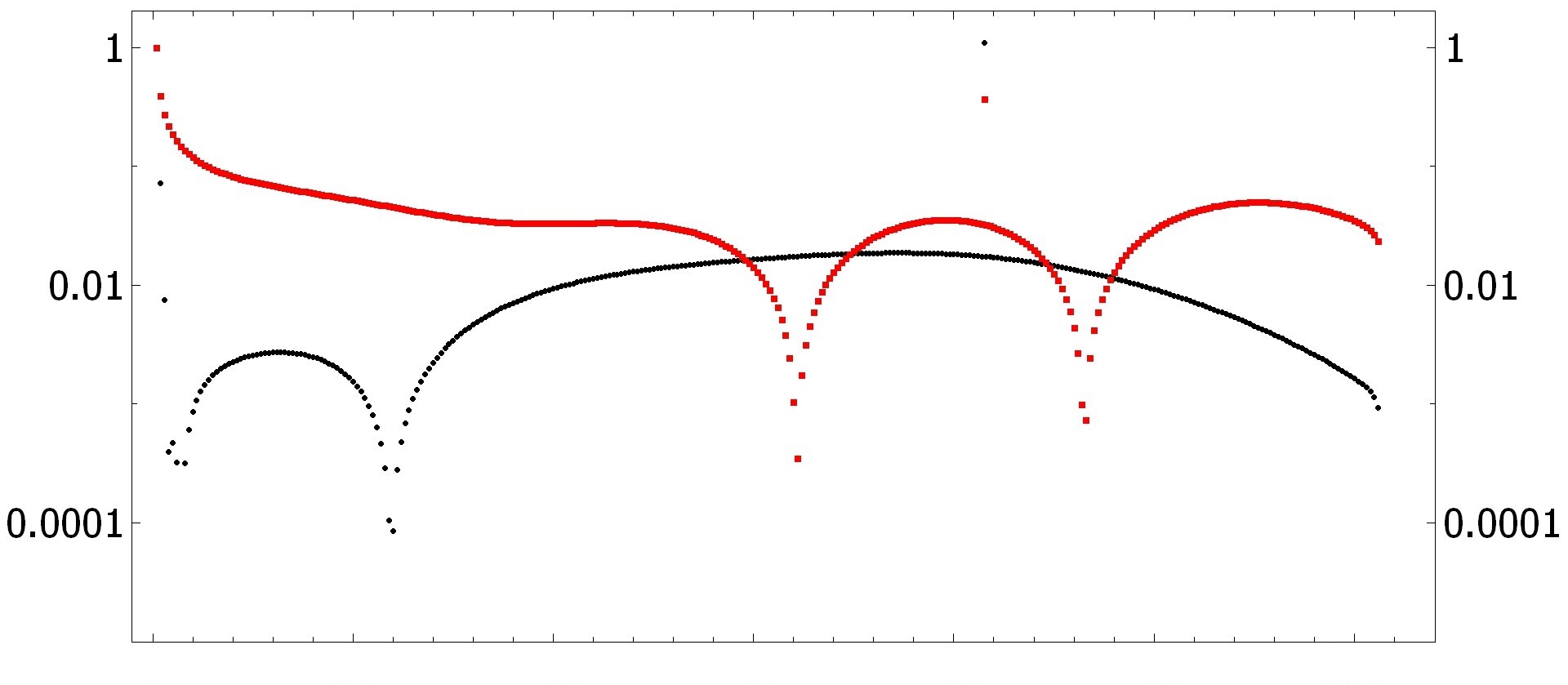}};
\node[below=of img1, node distance=0cm, yshift=1cm,font=\color{black}] {$3i-2$};
\node[left=of img1, node distance=0cm, rotate=90, anchor=center,yshift=-0.7cm,font=\color{black}] {Order parameter};
\node at (2.3, 2.1) {\tiny$|{\langle S^z_{3i-2}\rangle_{H+h_1 S^z_1 }}|$};
\node at (2.3, 1.7) {\tiny $|\langle Q^{zz}_{3i-2}\rangle_{H+q_1 Q^{zz}_1 }|$};
\node at (-3.4,-2.4) {$0$};
\node at (-2.3,-2.4) {$150$};
\node at (-1.2,-2.4) {$300$};
\node at (-0.1,-2.4) {$450$};
\node at (0.9,-2.4) {$600$};
\node at (2.05,-2.4) {$750$};
\node at (3.10,-2.4) {$900$};
\end{tikzpicture} \caption{Spin and SQ orders in the SQ phase, at $\theta=5^\circ$ and $J/K=0.1$, determined by measuring $S^z$ and $Q^{zz}$ on every third site after pinning $S^z$ and $Q^{zz}$ on the first site with large training fields $h_1=50$ and $q_1=50$, respectively. The system size is $L=918$. Flattening of the curves for a broad range of sites suggest a large correlation length.}
\label{SQ} 
\end{figure}
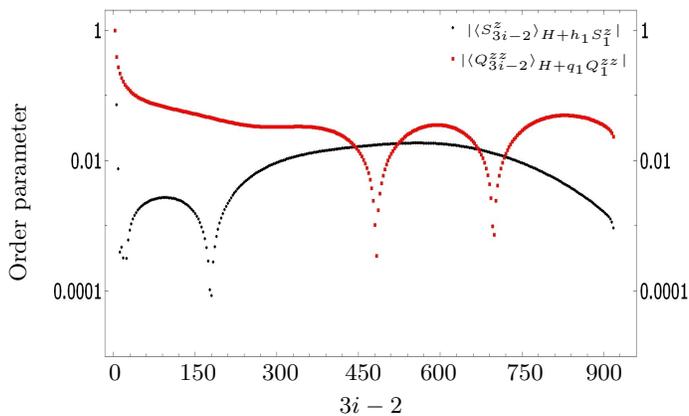

We apply pinning spin and SQ fields at the first site separately
and measure spin and SQ orders respectively far away from
pinning center and edges according to Eq. (\ref{eq:pinfield}). Again, $q=2\pi/3$ oscillations are removed by computing order parameters every three sites. The results are shown in Fig. \ref{SQ}. It shows that correlation length is extremely large and possibly diverges. Ref. \cite{nematic} shows that the correlation length indeed diverges at $J=0$, $0<\theta<\pi/4$ and the dominant correlations are SQ. At the systems sizes we can access, we are unable to determine decisively whether non-zero $J$ induces a gapped phase with exponentially decaying SQ correlations with a large correlation length or a critical phase like the $J=0$ limit. The resolution of this issue is left for future work.

\subsection{\label{sec:phase-transitions}Phase transitions out of the DSSCO}

\begin{figure}
\begin{tikzpicture}[black]
\node (img1)  {\includegraphics[height=5cm, width=8cm]{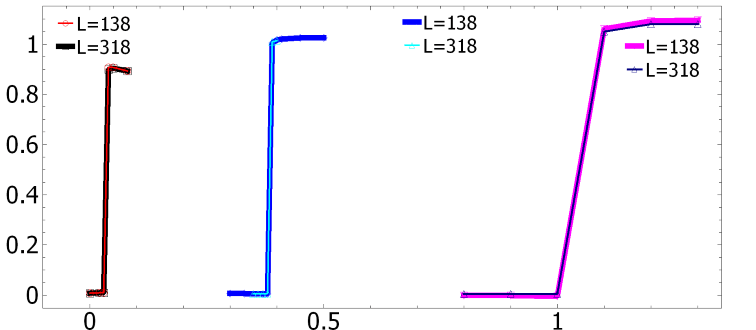}};
\node[below=of img1, node distance=0cm, yshift=1cm,font=\color{black}] {$J/K$};
\node[left=of img1, node distance=0cm, rotate=90, anchor=center,yshift=-0.7cm,font=\color{black}] {$\chi$};
\node[right=of img1, node distance=0cm,xshift=-1.5cm,yshift=1.05cm,font=\color{black}] {$\tiny0.8$};
\node[right=of img1, node distance=0cm,xshift=-1.5cm,yshift=1.85cm,font=\color{black}] {$\tiny1$};
\node[right=of img1, node distance=0cm,xshift=-1.5cm,yshift=0.25cm,font=\color{black}] {$\tiny0.6$};
\node[right=of img1, node distance=0cm,xshift=-1.5cm,yshift=-0.47cm,font=\color{black}] {$\tiny0.4$};
\node[right=of img1, node distance=0cm,xshift=-1.5cm,yshift=-1.25cm,font=\color{black}] {$\tiny0.2$};
\node[right=of img1, node distance=0cm,xshift=-1.5cm,yshift=-2.05cm,font=\color{black}] {$\tiny0$};
\node at (1.9, 2.1) { $\theta=24^\circ$};
\node at (-0.5, 2.1) {$\theta=15^\circ$};
\node at (-2.0, 2.0) {$\theta=2^\circ$};
\node at (3.0, 1.0) {DSSCO};
\node at (1.7, 1.0) {Disorder};
\node at (-0.5, 1.0) {DSSCO};
\node at (-1.8, 1.0) {Haldane};
\node at (-2.3, 0.0) {DSSCO};
\node at (-3.3, 0.0) {SQ};
\end{tikzpicture} \caption{Phase transitions out of the DSSCO phase into the disordered, Haldane and
SQ phases for two different sizes $L$. The abrupt change in $\chi$ as well as the weak $L$-dependence indicates
the first-order phase transitions. Here, uniform field $\lambda=0.01$ is used to train $\chi$}
\label{phasetran} 
\end{figure}
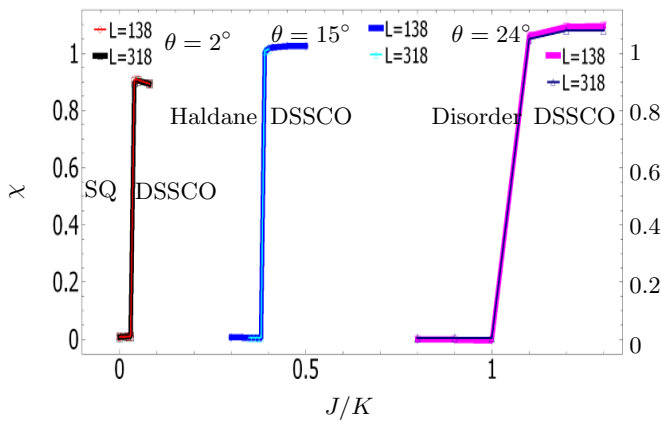

In Fig \ref{phasetran}, we show phase transitions from the DSSCO to the other phases which are Disordered phase, Haldane phase, and Spin-Quadrupole phase by calculating chiral order $\chi$ using Eq. (2) and (\ref{eq:2ndmeth}).
The abrupt drops in $\chi$ to zero clearly at the phase boundaries indicate first-order phase transitions.
Another indication of first-order transitions is the weak dependence
of $\chi$ on the system size. Since the correlation length does not diverge at the critical point for first order phase transitions, boundary effects are small, which result in a weak system-size dependence. Similar ideas were used to diagnose first order phase transitions in Ref. \cite{trimerization}.

\section{\label{sec:conclusions}Conclusions}

The DSSCO is a novel phase of matter that violates TRS
but has no density of moments, unlike other TRS-breaking phases known
in condensed matter. Using DMRG, we find that it appears when spin and
quadrupolar orders melt, leaving behind residual broken TRS but unbroken
continuous SRS. The chiral order is $O(1)$ in the DSSCO, which is much larger
than that in other existing chiral phases such as the chiral spin liquid, where $\chi\sim O(10^{-1})$ \cite{CSL}.
Besides, we propose a numerical method to study edge states by pinning
one edge and observing the other. It would be interesting to study other one-dimensional topological phases using this method.

\section*{Acknowledgements}

We would like to Miles Stoudenmire and Yixuan Huang for invaluable discussions. We acknowledge the Department of Physics and the College of Natural Sciences and Mathematics for financial support. We also acknowledge financial support from Chin-Sen Ting and Wu-Pei Su from Texas Center for Superconductivity at University of Houston .

 \bibliographystyle{apsrev4-1}
\bibliography{prbref}

\end{document}